\documentclass[preprint2]{aastex6}
\usepackage{apjfonts}
\usepackage{amsmath}

\begin{document}

\title{The quenched mass portion of star-forming galaxies and the origin of the star formation sequence slope}
\shortauthors{Pan et al.}
\shorttitle{The quenched mass of SFGs}
\author{Zhizheng Pan\altaffilmark{1}, Xianzhong Zheng \altaffilmark{1}, Xu Kong\altaffilmark{2}}

\altaffiltext{1}{Purple Mountain Observatory, Chinese Academy of Sciences, 2 West-Beijing Road, Nanjing 210008, China}
\altaffiltext{2}{CAS Key Laboratory for Research in Galaxies and Cosmology, Department of Astronomy, \\
University of Science and Technology of China, Hefei, Anhui 230026, China}
\email{panzz@pmo.ac.cn, xzzheng@pmo.ac.cn, xkong@ustc.edu.cn}

\begin{abstract}
Observationally, a massive disk galaxy can harbor a bulge component that is comparably inactive as a quiescent galaxy (QG). It has been speculated that the quenched component contained in star-forming galaxies (SFGs) is the reason why the star formation main sequence (MS) has a shallow slope at high masses. In this paper, we present a toy model to quantify the quenched mass portion of SFGs ($f_{\rm Q}$) at fixed stellar mass ($M_{\ast}$) and to reconcile the MS slopes both in the low and the high mass regimes. In this model, each SFG is composed by a star-forming plus a quenched component. The mass of the star-forming component ($M_{\rm SF}$) correlates with the star formation rate (SFR) following a relation SFR $\propto M_{\rm SF}^{\alpha_{\rm SF}}$, where $\alpha_{\rm SF}\sim 1.0$ . The quenched component contributes to the stellar mass but does not to the SFR. It is thus possible to quantify $f_{\rm Q}$ based on the departure of the observed MS slope $\alpha$ from $\alpha_{\rm SF}$. Adopting the redshift-dependent MS slope reported by \citet{Whitaker 2014}, we explore the evolution of the $f_{\rm Q}-M_{\ast}$ relations over $z=[0.5,2.5]$. We find that Milky-Way-like SFGs (with $M_{\ast}\approx 10^{10.7}M_{\sun}$) typically have a $f_{\rm Q}=30\%-40\%$ at $z\sim 2.25$, whereas this value rapidly rises up to $70\%-80\%$ at $z\sim 0.75$. The origin of an $\alpha\sim 1.0$ MS slope seen in the low mass regime is also discussed. We argue for a scenario in which the majority of low mass SFGs stay in a ``steady-stage" star formation phase. In this phase, the SFR is mainly regulated by stellar feedback and not significantly influenced by the quenching mechanisms, thus keeping roughly constant over cosmic time. This scenario successfully produces an $\alpha \sim 1.0$ MS slope, as well as the observed MS evolution from $z=2.5$ to $z=0$ at low masses.
\end{abstract}

\section{Introduction}\label{Sec1}

Large galaxy surveys have established that star-forming galaxies (SFGs) follow a relatively tight star formation rate (SFR)$-$stellar mass ($M_{\ast}$) relation from local universe to at least redshift $z\sim6.0$ \cite [e.g.,][]{Brinchmann 2004, Noeske 2007, Elbaz 2007, Karim 2011,Guo 2013,Speagle 2014,Tasca 2015}, with a dispersion of $\sigma\sim0.3$ dex in the logarithmic scale \citep{Guo 2013,Speagle 2014, Whitaker 2014,Kur 2016}. In a decade since its discovery, this star formation ``main sequence" (MS) is parameterized in a single power-law of SFR $=CM_{\ast}^{\alpha}$, in which $C$ and $\alpha$ are called the normalization parameter and the slope of the MS, respectively. Since the SFR$-M_{\ast}$ relation describes the stellar mass growth rate in galaxies at a given cosmic epoch, it has now been used as an important tool in studying galaxy assembly \cite[e.g.,][]{Leitner 2012,Patel 2013} and testing galaxy formation models \cite[e.g.,][]{Somerville 2015}.

In the past decade, much observational efforts have been focused on studying the SFR$-M_{\ast}$ relation at different cosmic epoches (see \citet{Speagle 2014} for a compilation of these works). The utilization of deep infrared photometry in SFR estimates enables a more robust characterization of this relation down to very low masses in the recent years. Using a mass-complete galaxy sample, \citet{Whitaker 2014} constrained the MS down to $M_{\ast}=10^{8.4}M_{\sun}$ ($M_{\ast}=10^{9.2}M_{\sun}$) at $z=0.5$ ($z=2.5$). For the first time, \citet{Whitaker 2014} reported that the MS is not consistent with a single power-law at $z=[0.5,2.5]$. Instead, it is better fitted in a broken power-law form, such that below a ``knee" mass ($M_{\rm k}$) of $M_{\rm k}=10^{10.2}M_{\sun}$, the MS has a redshift-independent slope of $\alpha\sim 1.0$. Above $M_{\rm k}$, the MS slope ranges from 0.2 to 0.7, depending on redshift. Subsequent works report a similar trend in the redshift range probed in \citet{Whitaker 2014}, but suggest that $M_{\rm k}$ may increase with redshift \citep{Lee 2015,Schreiber 2015,Tomczak 2016}. At higher redshifts ($z>3.5$), observations suggest that the MS also has a slope of $\alpha\sim 1.0$ at high masses \citep{Tasca 2015,Schreiber 2015}.

The shallow MS slope seen at $M_{\ast}>M_{\rm k}$ since $z\sim 2.5$ has been argued to be a consequence of the inclusion of quenched mass in massive galaxies \citep{Whitaker 2014,Lee 2015,Schreiber 2015,Erf 2016}. This idea can be easily interpreted since the quenched component of a SFG contributes to $M_{\ast}$ but does not to the SFR, thus naturally resulting in a flattening in the SFR$-M_{\ast}$ relation. The observations support this scenario. In the nearby Universe, it has been well established that the low mass SFGs ($M_{\ast}<10^{10.5}M_{\sun}$) are generally disk-dominated and composed by young stellar populations \cite[]{Kauffmann 2003}, whereas the more massive ones usually harbor a prominent bulge component containing relatively old stellar populations \citep{Abramson 2014, Gonzalez 2015}. More specifically, the specific star formation rate (sSFR) in the bulges of massive disk galaxies can be an order of magnitude lower than that in the disks, i.e., the bulges are indeed quenched \citep{Gonzalez 2016}. A similar picture appears to hold at higher redshifts. \citet{Nelson 2016} studied the stacked H$\alpha$ map of SFGs at $z=[0.7,1.5]$ and found that there is a strong central dip in the equivalent width of H$\alpha$ for massive SFGs, indicating the existence of a relatively old bulge component. Even at the cosmic star formation peak, there is evidence that the most massive SFGs ($\sim 10^{11.0}M_{\sun}$) have already harbored a quenched bulge component at $z\sim 2.2$, resembling their counterparts at lower redshifts \citep{Tacchella 2015}. Since the fraction of bulgy SFGs rapidly increases above $M_{\ast}\sim 10^{10.2}M_{\sun}$ \citep{Erf 2016, Pan 2016}, the quenched mass could be a potentially important contributor to the mass budget of SFGs.

The quenched mass portion contained in a SFG (hereafter referred as $f_{\rm Q}$) directly characterizes the ``maturity degree" of that galaxy, however is not well explored in previous studies. Obviously, $f_{\rm Q}$ can not be well constrained for individual galaxies without spatially-resolved information. However, one can possibly quantify the \emph{average} $f_{\rm Q}$ for the whole SFG population with very general information, as described below. In this paper, we develop a toy model to quantify the average $f_{\rm Q}$ at fixed $M_{\ast}$ by utilizing the bending effect of the quenched mass on the observed SFR$-M_{\ast}$ relation. The observations that less massive galaxies with $M_{\ast}<M_{\rm k}$ consistently follow a redshift-independent $\alpha \sim 1.0$ MS slope, lead us to the basic assumption of this model, that SFGs with a $f_{\rm Q}=0$ consistently follow a MS slope $\alpha_{\rm SF}$, where $\alpha_{\rm SF} \sim 1.0$. We justify this assumption and find that it is supported by the observations, as see below. Under this assumption, $f_{\rm Q}$ can be constrained by comparing the observed MS slope with $\alpha_{\rm SF}$. Throughout this paper, we adopt a concordance $\Lambda$CDM cosmology with $\Omega_{\rm m}=0.3$, $\Omega_{\rm \Lambda}=0.7$, $H_{\rm 0}=70$ $\rm km~s^{-1}$. Mpc$^{-1}$ and a \citet{Chabrier 2003} initial mass function (IMF).

\section{The model}\label{Sec2}
\begin{figure*}
\centering
\includegraphics[width=140mm,angle=0]{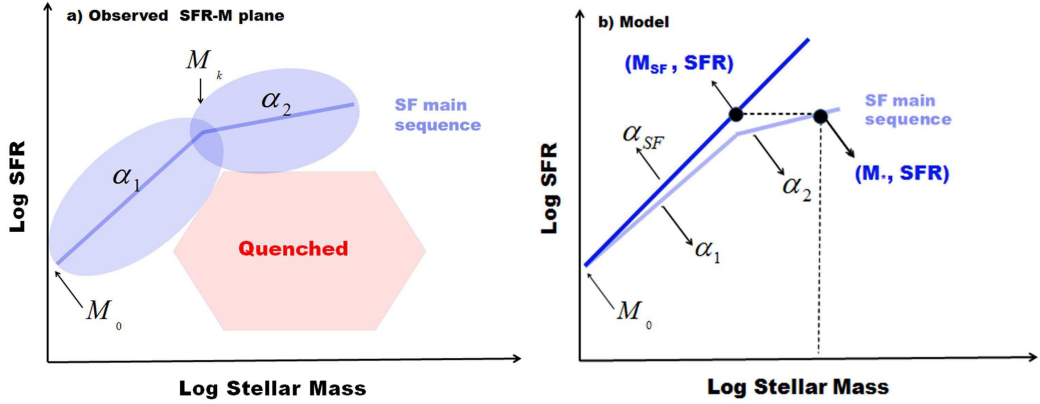}
\caption{a) A diagrammatic sketch of the observed log(SFR)$-$log($M_{\ast}$) plane. The SFGs lie on the tight MS, whereas the QGs locate in a much scattered region bellow the MS. The MS has an intrinsic dispersion of $\sigma \sim 0.3$ dex and can be described in a broken power-law form. Below a turn over mass $M_{k}$, the MS has a slope $\alpha_1 \sim 1.0$, while a shallower slope $\alpha_2$ is found above $M_{k}$. b) A diagrammatic sketch of our toy model. In this model, the SFR is only related to the star-forming component. }\label{fig1}
\end{figure*}

\citet{Whitaker 2014} for the first time claimed that there is a clear curvature in the MS at $z=[0.5,2.5]$. Therefore, the observed MS is better fitted with a broken power-law form
\begin{eqnarray}
 \mathrm{SFR(M_*)}=
\begin{cases}
C_1M_*^{\alpha_{1}},       & M_{*}\leq M_k \\
\\
C_2M_*^{\alpha_{2}},       & M_{*}\geq M_k \\
\end{cases}
\end{eqnarray}
where $M_{\rm k}\sim 10^{10.2}M_{\sun}$ and $\alpha_{\rm 1}$ and $\alpha_{\rm 2}$ are the slopes below and above $M_{\rm k}$, respectively. A diagrammatic sketch of the observed log(SFR)$-$log$(M_{\ast})$ plane is shown in the left panel of Figure~\ref{fig1}.

To reconcile the observed $\alpha_{1}$ and $\alpha_{2}$, we present a toy model in which a SFG is composed by a star-forming plus a quenched component, each following a SFR$-M$ relation
\begin{equation}
\mathrm{SFR(M_{SF}})=C_{\rm SF}M_{\rm SF}^{\alpha_{\rm SF}}, \mathrm{SFR(M_{Q}})=C_{\rm Q}M_{\rm Q}^{\alpha_{\rm Q}}
\end{equation}
where $M_{\rm SF}$ and $M_{\rm Q}$ are the stellar masses of the star-forming and quenched component, respectively. Regardless the exact form of the SFR$-M$ relation, the quenched component is expected to be only weakly correlated with the recent SFR. In observations, the $UVJ$-selected quiescent galaxies (QGs) \cite[e.g.,][]{Williams 2009} have average SFRs that are 20-40 times lower than SFGs, i.e., $C_{\rm Q}\ll C_{\rm SF}$. It is thus safely assuming that the observed SFR is primarily correlated with the star-forming component. Assuming the star-forming component accounts for a mass fraction of $f_{\rm SF}$, then
\begin{equation}
M_{\rm SF}=f_{\rm SF}M_{\ast}
\end{equation}
where $M_{\ast}$ is the total stellar mass of the galaxy. Given the total SFR is primarily correlated with the star-forming component, then
\begin{equation}
 \mathrm{SFR(M_\ast)}\approx\mathrm{SFR(M_{\rm SF})}=C_{\rm SF}f_{\rm SF}^{\alpha_{\rm SF}}M_{\ast}^{\alpha_{\rm SF}}
\end{equation}

\begin{figure*}
\centering
\includegraphics[width=140mm,angle=0]{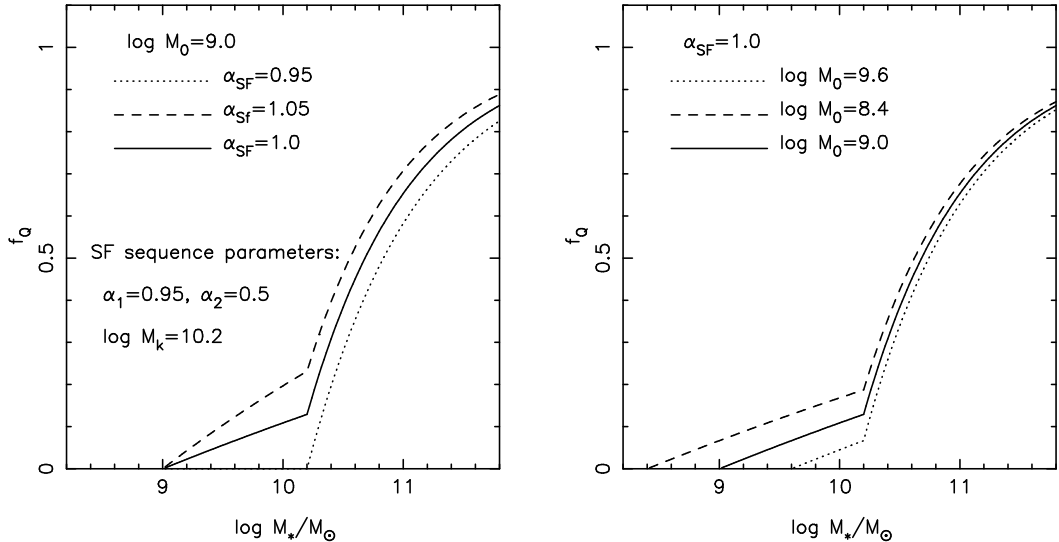}
\caption{Left:The dependence of the $f_{\rm Q}-M_{\ast}$ relation on $\alpha_{\rm SF}$. Right: The dependence of the $f_{\rm Q}$ relation on $M_{0}$. In each panel, we keep the MS parameters fixed.}\label{fig2}
\end{figure*}

Since the less massive SFGs are disk dominated and expected to contain little amount of quenched mass, then $\alpha_{\rm SF}$ should be similar to $\alpha_1$, i.e., $\alpha_{\rm SF}\approx \alpha_{1}$, as indicated in the right panel of Figure~\ref{fig1}. To eliminate the normalization parameters, one must take the boundary conditions into consideration. We assume that below a threshold mass of $M_{0}$, SFGs are solely composed by the star-forming component (see the right panel of Figure~\ref{fig1} ). Then at $M_{0}$
\begin{equation}
C_1M_{0}^{\alpha_1}=C_{\rm SF}M_{0}^{\alpha_{\rm SF}}
\end{equation}
Similarly, at $M_{\ast}=M_{\rm k}$
\begin{equation}
C_1M_{\rm k}^{\alpha_1}=C_2M_{\rm k}^{\alpha_2}
\end{equation}
Combining (1), (4), (5) and (6), then
\begin{eqnarray}
f_{\rm SF}=
\begin{cases}
\sqrt[\alpha_{\rm SF}]{M_0^{\alpha_{\rm SF}-\alpha_1}M_{\ast}^{\alpha_1-\alpha_{\rm SF}}},       &M_{0} \leq M_{\ast}\leq M_{k} \\
\\
\sqrt[\alpha_{\rm SF}]{M_0^{\alpha_{\rm SF}-\alpha_1}M_{k}^{\alpha_1-\alpha_2}M_{\ast}^{\alpha_2-\alpha_{\rm SF}}}       & M_{\ast}\geq M_k \\
\end{cases}
\end{eqnarray}
The quenched mass portion $f_{\rm Q}$ is then easily derived as $f_{\rm Q}$=1$-f_{\rm SF}$.

Note that $\alpha_1$, $\alpha_2$ and $M_{\rm k}$ can be determined from the MS, while $\alpha_{\rm SF}$ and $M_{\rm 0}$ are free parameters. As mentioned above, the observations suggest that $\alpha_{\rm SF}\sim 1.0$. In contrast, $M_{\rm 0}$ is not that well constrained. As defined, $M_{0}$ is a threshold mass below which the influence of quenching on galaxies could be ignored. In the local universe, \cite{Taylor 2015} found that the red population dissolves into obscurity when $M_{\ast}<10^{9.3}M_{\sun}$. A similar conclusion is also reached by \citet{Geha 2012}, who demonstrated that quenched galaxies with $M_{\ast}<10^{9.0}M_{\sun}$ do not exist in the field. Based on these studies, we suggest that $M_{\rm 0}$ should be smaller than $10^{9.5}M_{\sun}$.

To further explore the dependence of $f_{\rm Q}$ upon $\alpha_{\rm SF}$ and $M_{\rm 0}$, we show the $f_{\rm Q}-M_{\ast}$ relation as a function of $\alpha_{\rm SF}$ and $M_{0}$ in Figure~\ref{fig2}. As can be seen, the $f_{\rm Q}-M_{\ast}$ relation is not very sensitive to both $M_{0}$ and $\alpha_{\rm SF}$ in the mass regime of $M_{\ast}>M_{\rm k}$. This supports that the $f_{\rm Q}$ estimation is robust at high masses as long as the main sequence parameters are well determined. However, $f_{\rm Q}$ is strongly dependent on the choice of the free parameters at $M_{\ast}<M_{\rm k}$, implying that this method may be no longer valid in the low mass regime. We further discuss this in the caveat section.

It should be noted that there are non-ignorable uncertainties in the SFR and $M_{\ast}$ estimates, which may affect the measured MS parameters. Fortunately, these uncertainties only contribute to the dispersion of the MS relation and do not affect its slope \citep{Kur 2016}. Therefore, the estimated $f_{\rm Q}$ is robust against these measurement uncertainties. To conclude, this section presents a very simple model to quantify the $f_{\rm Q}-M_{\ast}$ relation of SFGs, which enables one to directly assess how ``mature" the galaxies are at a given $M_{\ast}$. The derived $f_{\rm Q}$ should be treated as an \emph{average} value since we have neglected the dispersion in the SFR$-M_{\ast}$ relation. Since $f_{\rm Q}$ is directly driven by the quenching processes, the evolution of the $f_{\rm Q}-M_{\ast}$ relation could be useful in constraining some key parameters of quenching, such as the average quenching rate and quenching time scale \citep{Lian 2016} in the future studies.

\section{A possible origin of the $\alpha_{SF}\sim 1.0$ slope}\label{Sec3}
Although the latest observational studies report a MS slope of $\alpha_1 \sim$1.0 at $M_{\ast}<M_{\rm k}$ \cite[e.g.,][]{Whitaker 2014, Schreiber 2015, Tomczak 2016}, the authors do not provide a plausible explanation to its origin. In this paper we argue that $\alpha_1\sim 1.0$ is totally expected since the low mass SFGs haven't underwent star formation suppression, as we will discuss bellow.

In principle, a SFG can grow its stellar mass via star formation or mergers. Specifically, the relative importance of these two channels depends on stellar mass and redshift. Both observations \cite[e.g.,][]{vanDokkum 2010,Own 2014,Vulcani 2016} and simulations \cite [e.g.,][]{De Lucia 2007,Guo 2008,Rod 2016,Qu 2016} suggested that mergers significantly contribute to the mass growth of massive galaxies ( $M_{\ast}\sim 10^{11.0}M_{\sun}$), especially at $z<1.0$. In contrast, the mass growth of low mass galaxies is dominated by in-situ star formation and the role of mergers is minor \citep{Leja 2015,Qu 2016}. As the analysis below is mainly focused on galaxies with $M_{\ast}<M_{\rm k}$, we will only consider the role of star formation in stellar mass growth and assume that mergers do not affect our final conclusion. Given this assumption, the stellar mass growth in a SFG between $t_0$ and $t_0+\Delta t$ is
\begin{equation}
 \Delta M=(1-R)\int_{t_0}^{t_0+\Delta t} \rm SFR(t)dt
\end{equation}
where $R$ is the return fraction due to mass loss. The return fraction $R$ is a function of time. However, as almost mass loss is within the first $10^8$ years, we fix $R$ to 0.36 (for a \citet{Chabrier 2003} IMF) in the following discussion.

The existence of a tight MS relation from at least $z \sim 6.0$ to $z=0$ suggests that the behavior of star-forming activity is predominately regular and smooth, rather than dominated by stochastic events like starbursts \cite[.e.g.,][]{Noeske 2007, Speagle 2014}. Specifically, the SFR can maintain at least over $\sim 10^{8}$ yr, as supported by the good consistency between the dust-corrected UV and H$\alpha$-based SFR, both in the local \citep{Hao 2011, k2012} and the high redshift universe \citep{Shivaei 2016}. In the case that the SFR keeps constant over the time interval probed, then equation (8) can be written as $\Delta M=\rm (1-R) SFR(t_0)\Delta t$.

Without quenching processes, then a SFG may maintain its SFR over a very long period after the SFR reaches a relatively stable value $\rm SFR_{\rm s}$. Assuming the SFR reaches $\rm SFR_{\rm s}$ at $t_s$ and keeps constant (or roughly constant) at $t>t_s$, then at any cosmic time $t$ (where $t>t_s$) its stellar mass $M(t)$ is
\begin{equation}
M(t)=M_s+\Delta M=M_s+\rm (1-R) SFR_s(t-t_s)
\end{equation}
where $M_{s}$ is the stellar mass formed prior to $t_s$. In the case of $(1-R){\rm SFR_s}(t-t_s)\gg M_s$, then $M(t)\approx (1-R)(t-t_s)\rm SFR_s$. For the whole SFG population, it should be reasonable assuming that the majority reaches a stable SFR at a similar $\overline{t_s}$. This assumption is supported by the simulation of \citet{Hopkins 2014}, who showed that galaxies reach their stable SFR at $z=3-6$. Therefore, the global SFR$-$$M_{\ast}$ relation can be expressed as log(SFR)=log($M_{\ast})$$-$log$(1-R)({t-\overline{t_s})}$ in the logarithmic space, which naturally results in an $\alpha\sim 1.0$ MS slope as observed.

\begin{figure*}
\centering
\includegraphics[width=160mm,angle=0]{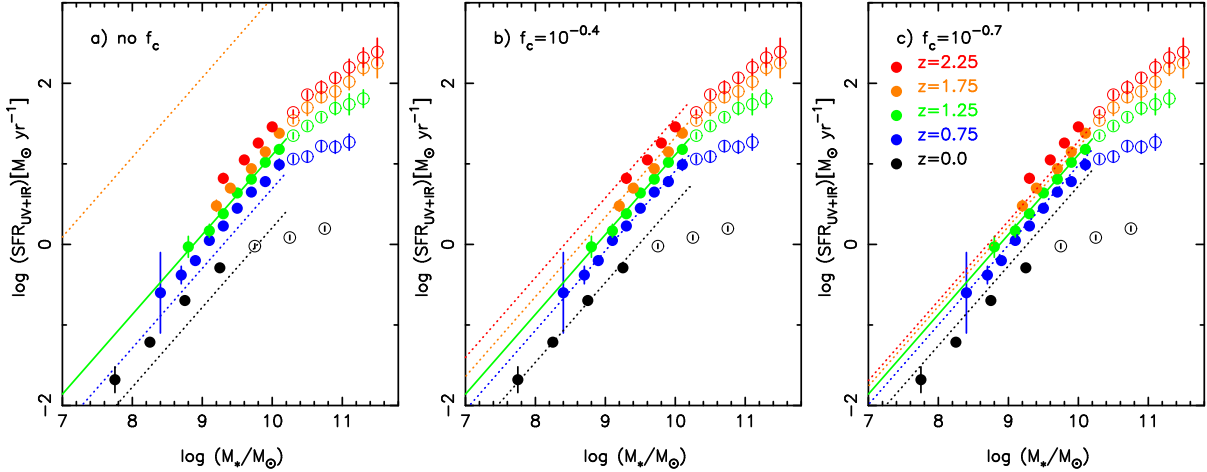}
\caption{Starting from the MS relation at $z=1.25$ (the green solid line), we predict the MS relations at $M_{\ast}<M_{\rm k}$ assuming that the SFRs are roughly unchanged over $z=[0,2.5]$. The observed MS data points at $z=[0.5,2.5]$ are from \citet{Whitaker 2014}, as indicated in the color symbols. The MS data at $z=0$ is from \citet{Gavazzi 2015}, as indicated in the black symbols. In each panel, the predicted MS relation at each redshift is shown in a color doted-line, with a same color-coding as the data points in that redshift. In panel a), the mass growth of SFGs is given by equation (8), with no additional $f_{\rm c}$ applied. In panel b) and c), the mass growth is given by equation (11). The solid symbols denote the data points that are used in the comparison (see the text for details).}\label{fig3}
\end{figure*}

However, this scenario should be over simple since it is not realistic for individual galaxies to strictly follow a constant SFR over a Hubble time. For SFGs, there are considerable variations in the star formation histories even at a fixed $M_{\ast}$, which are then manifested in the dispersion of the observed SFR$-M_{\ast}$ relation \citep{Cassara 2016}. Therefore, we emphasize that the ``constant SFR" assumption should be interpreted in a ``population-average" sense and does not necessarily hold for individual galaxies.

\section{The evolution of the main sequence at low masses}\label{Sec4}
In the above section we proposed a scenario in which SFGs keep their SFRs roughly constant over cosmic time. As a consequence, the stellar mass of a galaxy linearly correlates with cosmic time $t$, i.e., $M_{\ast}\propto \rm SFR\times t$, which then naturally results in an $\alpha \sim 1.0$ slope in the log(SFR)$-$log($M_{\ast}$) relation. This scenario predicts that the position of a SFG on the log(SFR)$-$log($M_{\ast}$) plane simply shifts towards higher masses as increasing $t$, while keeping its SFR unchanged. As such, given a log(SFR)$-$log($M_{\ast}$) relation at a starting cosmic time $t_{\rm start}$, one can predict the MS at any $t$. To verify this scenario, in this section we compare the model predicted MS relations with observations. We emphasize that this comparison is only meaningful for low mass galaxies with $M_{\ast}<M_{\rm k}$ since we assume that they are not significantly influenced by the quenching processes. The observed quenched fraction at $M_{\ast}<10^{9.5}M_{\sun}$ is quite low (<10\%) even in the local Universe \citep{Geha 2012, Taylor 2015}, supporting this assumption.

The observed MS data at $z=[0.5, 2.5]$ are drawn from \citet{Whitaker 2014}. We also complement the MS data at $z=0$ from \citet{Gavazzi 2015}.  The SFRs of \citet{Whitaker 2014} are estimated from the ultraviolet and 24 $\mu$m infrared photometry, while those of \citet{Gavazzi 2015} are from the observed H$\alpha$ fluxes. Both works have assumed a \citet{Chabrier 2003} IMF in the SFR and $M_{\ast}$ estimates. We choose a starting point of $z_{\rm start}=1.25$ to predict the MS relations at $z<1.25$ or at $z>1.25$ under the scenario proposed above. A $z_{\rm start}=1.25$ is chosen since at this redshift the observation reported MS slope ($\alpha_1=0.99$) is most close to our expectation. In fact changing $z_{\rm start}$ will not affect any of our conclusion. At $z=1.25$, \citet{Whitaker 2014} fit the MS with
\begin{equation}
\rm log~SFR=0.99\times(log~M_{\ast}-10.2)+1.31
\end{equation}
as shown in the green solid line in Figure~\ref{fig3}. Using the stellar growth given by equation (8), we predict the MS relations at four redshifts and show the results in the doted lines in panel a).  As can be seen, the predicted MS relations clearly do not match the observations. This may indicate that the proposed scenario is incorrect. However, we speculate that this disagreement may arise from the inconsistency between the cosmic SFR density and the stellar mass density growth rate ($\dot{\rho_{\ast}}$), as reported in some previous works \citep{Hopkins 2006,Wilkins 2008,Yu 2016}. These works found that the $\dot{\rho_{\ast}}$ inferred from the observed stellar mass functions is lower than the observed SFR density up to a factor of 0.2-0.6 dex. Taking this effect into consideration, we correct for the stellar mass growth due to the observed SFR by
 \begin{equation}
\Delta M=f_c \times (1-R)\rm {SFR_s}(t-t_0)
\end{equation}
where $f_{\rm c}$ is a correcting factor between 0.0 and 1.0.

We have tried a wide range of $f_{\rm c}$ to seek for a correcting factor that can result in a best matching between model predictions and observations. The explored $f_{\rm c}$ ranges from 0.0 to 1.0 dex, with a step of $\Delta f_{\rm c}=0.1$ dex. For each $f_{\rm c}$, the degree of the matching between model predictions and observations is then characterized by
\begin{equation}
\chi^{2} =\sum_{i=1}^{N}\frac{{(\rm SFR_{\rm i, predicted}-\rm SFR_{\rm i, observed})}^{2}}{N}
\end{equation}
where $\rm SFR_{\rm i, predicted}$ and $\rm SFR_{\rm i, observed}$ are the predicted and observed star formation rate, respectively. $N$ is the total number of data points that are considered for the comparison. Since we focus on galaxies with $M_{\ast}<M_{\rm k}$, thus only the data points with $M_{\ast}<10^{10.2}M_{\sun}$ are used for the comparison at $z=[0.5,2.5]$. At $z=0$, data points with $M_{\ast}<10^{9.5}M_{\sun}$ are used since the MS has turned over above this mass, as seen in Figure~\ref{fig3}. The data points that are used for comparison are denoted in solid symbols in Figure~\ref{fig3}. We find that a $f_{\rm c}=10^{-0.4}$ correcting factor yields the best matching (a minimal $\chi^{2}$=0.005) between model predictions and observations, as shown in panel b). This best-fit $f_{\rm c}$ value is well consistent with that reported in the previous studies. In panel c), it is clear that a $f_{\rm c}=10^{-0.7}$ correcting factor obviously underestimate the evolution of the MS.


It is still unclear why there is a systematic offset between the cosmic SFR density and $\dot{\rho_{\ast}}$. This may arise from the problems in stellar mass estimates, star formation rate estimates, or both (see \citet{Madau 2014} and \citet{Leja 2015} for a more detailed discussion). Nevertheless, the good consistency between the model predicted and observed MS relations over a wide redshift range still strongly support this scenario.

\section{The evolution of $f_{Q}$ and quenched mass density over z=[0.5,2.5]}\label{Sec5}
In this section we adopt the MS parameters given by \cite{Whitaker 2014} to explore the evolution of $f_{\rm Q}$ over $z=[0.5, 2.5]$. \cite{Whitaker 2014} fit a broken power law form for the MS relation with a fixed turn over mass of $M_{\rm k}=10^{10.2}M_{\sun}$. The slope is close to unity at $M_{\ast}<M_{\rm k}$, for which \cite{Whitaker 2014} give
\begin{equation}
\alpha_1(z)=0.95\pm0.05+(0.02\pm0.04)z
\end{equation}
where $z$ is redshift. Above $M_{k}$, the slope is strongly redshift-dependent, with a form of
\begin{equation}
\alpha_2(z)=0.03\pm0.10+(0.31\pm0.06)z
\end{equation}

\begin{figure}
\centering
\includegraphics[width=80mm,angle=0]{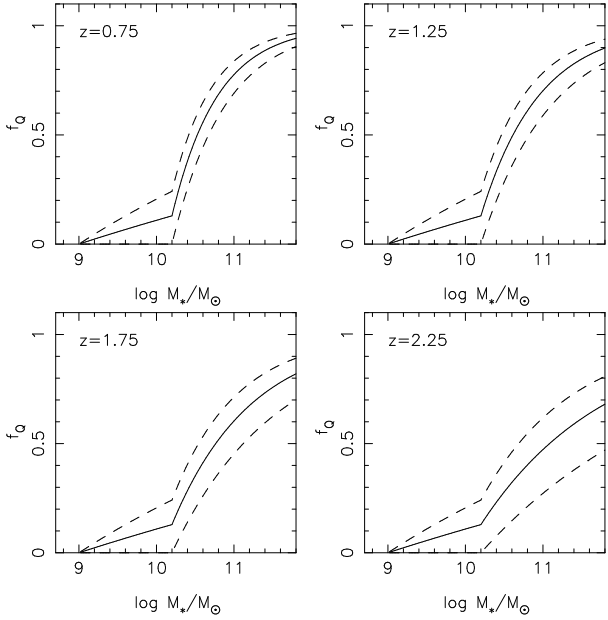}
\caption{The solid lines show the $f_{\rm Q}-M_{\ast}$ relations at $z=[0.5,2.5]$, adopting the $M_{k}$, $\alpha_1$, $\alpha_2$ reported by \citet{Whitaker 2014}. The dashed lines indicate the upper and lower limits of $f_{\rm Q}$ by taking the uncertainties in $\alpha_1$ and $\alpha_2$ into consideration. As reported in \citet{Whitaker 2014}, the uncertainties are $\Delta \alpha_1=\pm 0.05$ and $\Delta \alpha_2=\pm 0.1$, respectively. }\label{fig4}
\end{figure}

To simplify, we adopt $\alpha_{\rm SF}=1.0$ and $M_{0}=10^{9.0}M_{\sun}$ in this section. The derived $f_{\rm Q}-M_{\ast}$ relations are shown in Figure~\ref{fig4}. As can be seen, the $f_{\rm Q}$ of a Milky-Way-like SFG (with $M_{\ast}\approx10^{10.7}M_{\sun}$) is around $30\%-40\%$ at $z\sim 2.25$, whereas it rapidly rises up to $70\%-80\%$ at $z\sim 0.75$. This indicates that the massive SFGs have been dominated by quenched mass since very high redshifts.


\begin{figure*}
\centering
\includegraphics[width=120mm,angle=0]{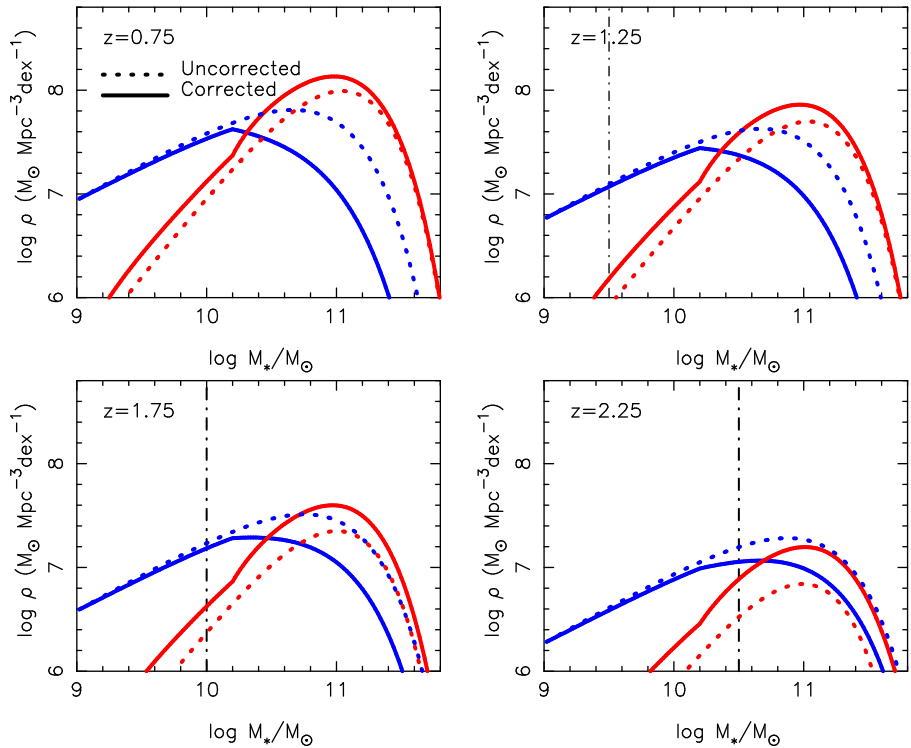}
\caption{The sellar mass density distributions for the star-forming and quenched component at different redshifts, as indicated by the blue and red lines, respectively. The doted lines are derived from the stellar mass functions of \citet{Muzzin 2013}. The solid lines show the corrected stellar mass density distributions given in equation (15) and (16). The vertical dot-dashed lines show the limited stellar masses of the stellar mass functions.}\label{fig5}
\end{figure*}

An important feature of Figure~\ref{fig4} is that even at $z=2.25$, the most massive SFGs have already contained a considerably high fraction of quenched mass. For example, the $f_{\rm Q}$ can be greater than 50\% for SFGs with $M_{\ast}>10^{11.3}M_{\sun}$. This is broadly consistent with the findings of \citet{Tacchella 2015}.  \citet{Tacchella 2015} reach their conclusions based on a small sample of 22 SFGs (only 5 with $M_{\ast}>10^{10.7}M_{\sun}$), which may be seriously biased to selection effects. Since the $f_{\rm Q}$ estimation depends on the parameters of the MS, Figure~\ref{fig4} thus confirms their finding in a more statistical sense.

Using the stellar mass functions, early works have estimated star-forming and quenched stellar mass density in the universe \citep{Ilbert 2013, Muzzin 2013, Tomczak 2014}. These works do not consider the hidden quenched mass in SFGs, thus will certainly underestimate the true quenched mass density. We revisit this issue by taking $f_{\rm Q}$ into consideration. The corrected mass density for the star-forming and quenched component should be
\begin{eqnarray}
\rho_{\rm SF,corr}=(1-f_{\rm Q})\rho_{\rm SF, SMF}\\
\rho_{\rm Q,corr}=\rho_{\rm Q, SMF}+f_{\rm Q}\rho_{\rm SF, SMF}
\end{eqnarray}
where $\rho_{\rm Q, SMF}$ and $\rho_{\rm SF, SMF}$ are the stellar mass density of QGs and SFGs derived from their stellar mass functions, respectively. In this paper we adopt the stellar mass functions reported by \citet{Muzzin 2013}, who also use the $UVJ$ technology to define SFGs and QGs. The results are shown in Figure~\ref{fig5}. As can be seen, without correcting for $f_{\rm Q}$, the quenched mass density takes over at $z<1$, which is already reported in early works. Once $f_{\rm Q}$ is taken into consideration, the quenched mass is already taking over at $z\sim2.25$ for galaxies more massive than $10^{10.7}M_{\sun}$. The results of Figure~\ref{fig5} are broadly consistent with the recent findings of \citet{Renzini 2016}.

Figure~\ref{fig5} reveals that the quenched mass density doubles (increase $\sim 0.3$ dex) after correcting for $f_{\rm Q}$ at $z=2.25$, i.e., the SFGs contribute equally or even more than QGs do to the total quenched mass budget. Although remains to be confirmed, this could be the case in the early cosmic epoch when quenching just started at work. As the quenching processes proceed in an inside-out manner in massive SFGs \cite [e.g.,][]{Pan 2014, Li 2015,Tacchella 2015, Pan 2015, Pan 2016, Belfiore 2016}, quenched mass have already emerged in the bulge of a SFG prior to the fully quenching of that galaxy. Therefore, it makes sense that the quenched mass primarily exists in the bulges of massive SFGs when quenching is still at its preliminary stage.

\section{Discussion}
We have developed a toy model to interpret the slope of the log(SFR)$-$log($M_{\ast}$) relation of star-forming galaxies. Our model splits a SFG into a quenched plus a star-forming component, which is initially inspired by the observations that bulges are generally quenched while disks are still forming stars \citep{Abramson 2014}. However, we emphasize that the quenched mass mentioned in this paper is not strictly equal to the bulge mass for two reasons. First, star-forming bulges do also exist, especially at high redshifts \citep{Barro 2013, Whitaker 2015}. Second, a considerable fraction of the old stars can migrate from the bulge to the outer disk due to resonant scattering with transient spiral arms \citep{Sellwood 2002,Rovkar 2008,Loebman 2016, Badry 2016}, making these two components indistinguishable even with the help of a disk-bulge decomposition.

In Section~\ref{Sec3} we argue that the MS will naturally have an $\alpha \sim 1.0$ slope once SFGs enter a steady-stage star formation phase. Section~\ref{Sec4} further verifies the capability of this scenario in explaining the observed evolution of the MS over $z=[0.0, 2.5]$ at $M_{\ast}<M_{\rm k}$. In fact the notion that low mass galaxies generally have an extended star formation history is not new. Previous works on galaxy stellar populations have indicated that while the massive galaxies form the majority of their stellar mass at $z\sim 2-3$, the less massive ones grow their mass with a similar speed over a Hubble time \cite [.e.g.,][]{Thomas 2005, McDermid 2015}. A similar conclusion is also reached by studies with independent approaches. For example, \citet{Behroozi 2013} and \citet{Moster 2013} study the average star formation history of galaxies in dark matter haloes from high redshifts to the present day using an abundance matching method. They both find that the low mass haloes ($M_{h}<10^{12}M_{\sun}$) have a steady SFR since $z\sim 2.0$.

This steady-stage star formation phase is likely a consequence of stellar feedback. Stellar feedback has long been served as an important mechanism that shaping various properties of galaxies in the low mass regime, including the stellar surface density \citep{Kauffmann 2003}, the metallicity \citep{Tremonti 2004} and the morphologies \citep{Brook 2011}. The star-of-the-art numerical simulations have successfully produced a low star formation efficiency in low mass haloes as seen in observations by taking explicit stellar feedback physics into consideration \cite[.e.g.,][]{Hopkins 2014, Ceverino 2014}. In the FIRE simulation, a galaxy will reach a steady SFR phase at $z\sim 3-6$, at which the stellar feedback appears to dominate gas dynamics \citep{Hopkins 2014}.  Without additional quenching mechanisms (such as active galactic nucleus (AGN) feedback or halo shock heating) involved, the simulations indicate that the galaxy will maintain its SFR for a long period (see Figure 10 of \citet{Hopkins 2014}).

We also note that a $f_{\rm c}\sim 10^{-0.4}$ correcting factor is useful in interpreting some recent findings. \citet{Tomczak 2016} used the MS relation to predict the evolution of stellar mass function from $z=2.5$ to $z=0.5$. They found that the galaxy number density $\Phi(M_{\ast})$ is systematically over predicted by $\sim$ 0.2 dex at $M_{\ast}<M_{\rm k}$ (see their Figure 10). Since the stellar mass function has a faint end slope of $\alpha \sim -1.5$, this effect can be equivalently interpreted as a $\sim$ 0.4 dex overestimation in $M_{\ast}$, as \citet{Tomczak 2016} did not include a $f_{\rm c}$ factor in their calculations.

In Section~\ref{Sec5} we show that $f_{\rm Q}$ is already considerably high for massive SFGs at the peak of cosmic star formation ($z\sim 1.5-2.5$). As the quenched mass is primarily associated with bulges, this indicates that the bulge buildup process is highly efficient at high redshifts. At $z\sim 2.0$, galaxies are gas rich and typically have a gas fraction around 0.4-0.5 \citep{Tacconi 2010,Tacconi 2013}. With such a high cold gas fraction, a large amount of the gas will sink into the centers of galaxies due to violent disc instability, triggering central starbursts and forming a prominent bulge \citep{Dekel 2014}. This merger-free bulge forming scenario is supported by the recent work of \citet{Tadaki 2016}, who use ALMA and KMOS observations on 25 main sequence galaxies to reveal that the rotation-supported SFGs at $z\sim 2.0$ have very intense central star formation rate. They conclude that these galaxies are able to form a compact bulge with a central 1kpc stellar mass density $\Sigma_{M_{\ast},1kpc}>10^{10}M_{\sun}~\rm kpc^{-2}$ in a few $10^8$ years.

The link between the emergence of a quenched bulge and the fully quenching of a massive galaxy is still missing. Bulges are generally quenched, indicating that quenching may first operate from the inner galaxy regions. This is likely associated with a central starburst episode \citep{Tacchella 2015, Tacchella 2016,Tadaki 2016}, which is possibly accompanied by strong gas outflow driven by AGNs \citep{Genzel 2014}. However, an external process such as the shut-down of cold gas accretion may also needed to explain the subsequent suppression of star formation in the outer part of the galaxy \citep{Dekel 2006}. The detailed quenching mechanism is key to galaxy evolution \cite[e.g.,][]{Peng 2010}, however, is beyond the scope of this paper.

\section{Caveat}
Several issues to the analysis presented above warrant some considerations.First, the analysis presented in section~\ref{Sec3} and section~\ref{Sec4} suggests that $\alpha_{\rm SF}\sim 1.0$ is indeed the case at $M_{\ast}<M_{\rm k}$. However, whether $\alpha_{\rm SF}\sim 1.0$ holds at $M_{\ast}>M_{\rm k}$ is not well justified. Note that in section~\ref{Sec3} we have ignored mergers. At high masses, the role of mergers in mass growth can not be easily ignored. In merger remnants, it is not clear whether the star-forming component still follow a same SFR$-M_{\rm SF}$ relation as their progenitors. However, if the properties of the star-forming component are not significantly changed during the merging process, the scaling relation between SFR and $M_{\rm SF}$ should remain for the merger remnants.

Second, the $f_{\rm Q}$ derived in this model should be treated as an \emph{upper limit} of the true value since we have assumed that the quenched mass is fully responsible for the flattening of the MS. This is not well justified and there may also exist other mechanisms that can lead to the flattening. For example, \citet{Schreiber 2016} claimed that the massive SFGs at $z\sim 1.0$ have a decreased star formation efficiency that can up to a factor of 3 compared to the less massive ones, thus responsible for the flattening of the MS. However, the findings of \citet{Schreiber 2016} is not seen in the local Universe. Recently, \citet{Saintonge 2016} explored the mean atomic and molecular gas mass fraction along the MS at $z=0$ with the data from the ALFALFA, GASS and COLD GASS surveys. In contrary to \citet{Schreiber 2016}, \citet{Saintonge 2016} found that both star formation efficiency and molecular-to-atomic gas ratio vary little for massive SFGs, indicating the flattening of the local MS is due to the global decrease of the cold gas reservoir rather than a depression in star formation efficiency.

Third, the model may be less physical for less massive SFGs. This is because less massive galaxies are disk dominated and do not harbor a notable quenched component. In addition, the model predicted $f_{\rm Q}$ is strongly dependent on $M_{0}$ and $\alpha_{\rm SF}$ at low masses (see Figure~\ref{fig2}), both of which are free parameters that can not be firmly constrained. Although the model predicts a very low $f_{\rm Q}$ in the low mass regime that still seems reasonable, we suggest that this value is not meaningful.

\section{Summary}
In this paper, we develop a toy model to quantify the quenched mass portion ($f_{\rm Q}$) of SFGs and to reconcile the star formation sequence slopes both in the low and the high mass regimes. Our results are summarized as follows.

1. In this toy model, each SFG is composed by a star-forming plus a quenched component. The mass of the star-forming component ($M_{\rm SF}$) correlates with the SFR following a relation SFR $\propto M_{\rm SF}^{\alpha_{\rm SF}}$, where $\alpha_{\rm SF}\sim 1.0$. The quenched component contributes to the total stellar mass while doesn't to the SFR, thus driving the observed MS slope $\alpha$ depart from $\alpha_{\rm SF}$. The difference between $\alpha_{\rm SF}$ and $\alpha$ thus can be used to infer $f_{\rm Q}$, as given in equation (7).

2. We propose a scenario to interpret the origin of the $\alpha\sim 1.0$ main sequence slope seen at $M_{\ast}<M_{\rm k}$. In this scenario, the majority of low mass SFGs are less influenced by the quenching processes, thus keeping a steady SFR over cosmic time. As such, the stellar mass of a galaxy linearly correlates with cosmic time $t$, i.e., $M_{\ast}\propto \rm SFR\times t$, which then naturally results in an $\alpha \sim 1.0$ slope in the log(SFR)$-$log($M_{\ast}$) relation.  The observed MS relations in the low mass regime agree well with the model predictions at $z=[0,2.5]$, which supports this scenario. We suggest that the steady-stage star formation phase is driven by stellar feedback.

3. Adopting the redshift-dependent main sequence slope reported by \citet{Whitaker 2014}, we explore the evolution of the $f_{\rm Q}-M_{\ast}$ relation over $z=[0.5,2.5]$. We find that Milky-Way-like SFGs typically have a $f_{\rm Q}=30\%-40\%$ at $z\sim 2.25$, whereas this value rapidly rises up to $70\%-80\%$ at $z\sim 0.75$. Taking $f_{\rm Q}$ into consideration, we retrieve the stellar mass density budget of the universe, finding the quenched mass has been taking over in galaxies with $M_{\ast}>10^{10.7}M_{\sun}$ since $z\sim 2.25$.

\acknowledgments
We thank the anonymous referee for a very constructive report that helped to improve the quality of this paper. This work is supported by the Chinese National 973 Fundamental Science Programs (973 program) (No 2015CB857004).

\end{document}